\documentclass[review]{elsarticle}

\usepackage{amssymb}
\usepackage{url}
\usepackage{amsfonts,amsmath}
\usepackage{graphicx}
\usepackage{stfloats}
\usepackage{subfigure}

\usepackage{multirow}
\usepackage{xfrac}
\usepackage{color}
\usepackage{array}
\usepackage[table]{xcolor}

\usepackage[Symbol]{upgreek}

\usepackage[T1]{fontenc}

\usepackage{graphicx,url}
\usepackage{picinpar}
\usepackage{amsmath}
\usepackage{stfloats}
\usepackage{url}
\usepackage{flushend}
\usepackage[latin1]{inputenc}
\usepackage{colortbl}
\usepackage{soul}
\usepackage{multirow}
\usepackage{pifont}
\usepackage{color}
\usepackage{alltt}
\usepackage[hidelinks]{hyperref}
\usepackage{enumerate}
\usepackage{siunitx}
\usepackage{breakurl}
\usepackage{epstopdf}
\usepackage{pbox}
\usepackage{algorithm}
\usepackage{algpseudocode}

\journal{arXiv.org}

\biboptions{authoryear}

\begin{document}

\begin{frontmatter}

\title{Application-Specific System Processor for the SHA-1 Hash Algorithm}


\author[UFRN]{Carlos~E.~B.~S.~J\'unior}
\ead{ceduardobsantos@gmail.com}

\author[SU]{Matheus~F.~Torquato}
\ead{m.f.torquato@swansea.ac.uk}

\author[UFRN]{Marcelo~A.~C.~Fernandes\corref{cor1}}
\ead{mfernandes@dca.ufrn.br}

\cortext[cor1]{Corresponding author}
\address[UFRN]{Department of Computer Engineering and Automation, Federal University of Rio Grande do Norte (UFRN), Natal, Brazil}
\address[SU]{College of Engineering, Swansea University, Swansea, Wales, UK}





\begin{abstract}
This work proposes an Application-Specific System Processor (ASSP)  hardware for the Secure Hash Algorithm 1 (SHA-1) algorithm. The proposed hardware was implemented in a Field Programmable Gate Array (FPGA) Xilinx Virtex 6 xc6vlx240t-1ff1156. The throughput and the occupied area were analyzed for several implementations in parallel instances of the hash algorithm. The results showed that the hardware proposed for the SHA-1 achieved a throughput of 0.644 Gbps for a single instance and slightly more than 28 Gbps for 48 instances in a single FPGA. Various applications such as password recovery, password validation, and high volume data integrity checking can be performed efficiently and quickly with an ASSP for SHA1.
\end{abstract}


\end{frontmatter}


\section{Introduction}\label{sec:intro}

The Secure Hash Algorithm version one, SHA-1, is an algorithm used to verify the integrity of variable length data streams from an operation called hash. A hash function outputs a fixed-length code $C$ given a message of variable length $K$ as input. It may be said that the output of the hash function, also called the hash code, is a signature of the input message, known in the literature as a fingerprint. These features are used in Message Authentication Codes (MAC), mainly the Keyed-Hash Message Authentication Code (HMAC), which make extensive use of SHA-1 \citep{Ref3SHA1}.

The SHA-1 is a revised version of the SHA-0, a substitute for the MD5 (Message Digest 5) algorithm in 1995 by the National Institute of Standards and Technology (NIST). Then, the SHA-1 was published as a Federal Information Processing Standard (FIPS) number 180-1 \citep{stallings:15}. 

The SHA-1 hash function, in addition of being selected for Digital Signature Algorithm (DSA) as standardized by FIPS 186-4 \citep{dss:13}, it acts verifying the sequence of data associated with communication protocol messages, files, passwords storage and it was used in digital certificates before being replaced by SHA-2. The SHA-1 generates hash codes of $C = 160 $ bits for any size of $ K $. However, with the advent of the Big Data, Internet of Things (IoT) and other emergent areas. It is necessary to the hash code to be generated more quickly for situations associated with a large volume of data and with small energy consumption in the case of devices within a sensor network in the IoT context. Thus, this project aims to develop a dedicated hardware implementation proposal for the SHA-1 algorithm. The proposed hardware can be seen as a specific application processor also called in the literature by Application-Specific System Processor (ASSP).

The presented hardware was developed in a Field-programmable gate array (FPGA), reconfigurable hardware platform formed by thousands of logical cells, that after a synthesis process behaves as specific hardware associated to a given algorithm. The FPGA has been an indispensable tool in the development of ASSPs, Application Specific Integrated Circuits (ASICs) and as a platform for accelerating complex algorithms as presented in \citep{Ref10,Ref11,Ref12,Ref13,RefLivro1}. One of the advantages in developing specific circuits is the clock reduction when compared to implementations in systems with general purpose processors (GPP). The hardware SHA-1 algorithm can be used in the development of an ASIC for IoT applications or used in the FPGA itself, aiming to accelerate hash code calculation in several applications such as password recovery, password validation and integrity checking in large volumes of data.

\section{Related Work} \label{sec:trabalhos}

The work presented in \cite{Ref1SHA1} used a Xilinx Virtex-II XC2V2000-6 FPGA to implement the SHA-1 with Iterative Looping (IL). This implementation occupied around $ 1,275 $ Logic Cells (LC) operating at a throughput of 734Mbps. The works presented in \cite {Ref2SHA1,Ref3SHA1} also used a Xilinx Virtex-II XC2V2000-6 FPGA to implement the SHA-1. These implementation proposals present a scheme using Full Pipeline (FL) that consumed around $3,519 $ LC for a throughput of $2.5267$ Gbps. Comparing with the proposal presented in \cite {Ref1SHA1}, the throughput is $ 4 \times$ higher due to the use of $ 4 $ SHA-1 pipelined modules. However, the occupancy area is also around $4 \times$ larger. In \cite{Ref3aSHA1} it is also presented a proposal using FL that reached a throughput of $5.9$ Gbps.

In \cite{Ref4SHA1} the SHA-1 implementation was executed in a Xilinx Virtex 5 Xc5vlx50t FPGA using hardware description language {Verilog}. The implementation performed with IL was similar to that presented in \cite{Ref1SHA1}. However, it had a slightly higher occupancy rate, around 1,351 LC, and the throughput also a little higher, around $786$ Mbps.

The work presented by \cite{Ref5SHA1} brought a solution of the SHA-1 in FPGA with low power for uses in devices lacking high power capacity and with high throughput, in addition to a small area size compared to the similar implementations. For this, the authors relied on the work presented in \cite {Ref2SHA1} and \cite{Ref3SHA1}. In this work the LC number was reduced by making a more serial implementation, also reducing the throughput. Another implementation-based approach described in \cite {Ref2SHA1} and \cite{Ref3SHA1} was showed in \cite{Ref6SHA1}, in which an implementation in a TSMC 90 nm ASIC was proposed. In this proposal, a throughput of around $15$ Gbps was observed.

A comparison of several Xilinx FPGA platforms with SHA-1 implementation was presented in \cite {Ref7SHA1}. The implementation was based on the proposal presented in \cite {Ref2SHA1} and \cite {Ref3SHA1} and a maximum throughput of about $14.3$ Gbps was observed for a Xilinx Virtex 7 FPGA.

Works with SHA-1 implementation on other hardware platforms can be found in \cite{Ref1GPUSHA1} and \cite{Ref2GPUSHA1} in which comparisons between Graphics Processing Units (GPUs) and CPUs were performed. The GPUs NVIDIA Tesla M2050 with $448$ CUDA cores and  AMD FirePro V7800 with  $1440$ stream processors could achieve throughput peaks of up to $1.5$ Gbps.

The proposal here developed used as target device a Virtex FPGA 6 xc6vlx240t-11156 FPGA and the results showed a throughput of $652$ Mbps for a single SHA-1 module. The implementation used the Iterative Looping strategy which occupied less circuit area when compared to other strategy \cite {Ref2SHA1} and \cite{ Ref3SHA1} and unlike the results presented in the literature, it was possible to synthesize up to $48$ SHA-1 modules in a single FPGA device yielding a throughput of $28.160$ Gbps.

\section{Secure Hash Algorithm 1 (SHA-1)}\label{sec:algoritmo}

The SHA-1 is a hashing algorithm described by the Federal Information Processing Standards Publication (FIPS) 180-4 \cite {fips} and by RFC 3174 \cite {rfc}, which operates with variable length input messages. For each $i$-th incoming message, $m_i$, (of length $K_i$ bits), expressed as
\begin{equation}
\mathbf{}{m}_i = \left[
\begin{array}{cccc}
m_{0} & m_{1} & \dots & m_{K_i-1}
\end{array}
\right] \textnormal{ where } m_{k} \, \in \, \{0,1\} \, \forall \, k,
\label{eq1}
\end{equation}
the SHA-1 algorithm generates an output message, $m_i$, called a hash code, of fixed size $ C = 160 $ bits, characterized as
\begin{equation}
\mathbf{}{h}_i = \left[
\begin{array}{cccc}
h_{0} & h_{1} & \dots & h_{C-1}
\end{array}~
\right] \textnormal{ where } h_{k} \, \in \, \{0,1\} \, \forall \, k.
\label{eq2}
\end{equation}

The $i$-th incoming message, $m_i$, of $K_i$ bits is extended by inserting two binary words. The first one, called here, $p_i$, has $P_i$ bits and it is inserted by an operation called  Append Padding. The second, called here $v_i$, has $T$ bits and it is inserted by an operation called Append Length. Thus, the calculation of the hash code, $h _i$, for each $i$-th incoming message is carried out in an extended message, here called $z_i$, which corresponds to a concatenation of the messages $m_i$, $p_i$ and $v_i$, that is, $z_i = [m_i , p_i v_i]$. Each $i$-th message $z_i$ has $Z_i = K_i + P_i + T $ bits that can be divided into $L_i$ blocks of length $ M = 512 $ bits, that is,
\begin{equation}
L_i = \frac{Z_i}{M} = \frac{K_i+P_i+T}{512}.
\label{eq2a}
\end{equation}

The pseudo-code presented in the Algorithm \ref{ResumoSHA} displays the sequence of steps required to generate the hash code. These steps are going to be described in detail in the following subsections.

\begin{algorithm}
	\caption{SHA-1 for each $i$-th message $W_i$}\label{ResumoSHA}
	\begin{algorithmic}[1]
		\State $\mathbf{z}_i \gets  [\mathbf{m}_i]$ \label{Cod1}
		\State $\mathbf{p}_i \gets \text{{PaddingGeneration}}(K_i)$ \label{CodGP}
		\State $\mathbf{z}_i \gets  [\mathbf{m}_i \, \mathbf{p}_i]$ \label{CodIP}
		\State $\mathbf{v}_i \gets \text{{LenghtGeneration}}(K_i)$ \label{CodGC}
		\State $\mathbf{z}_i \gets  [\mathbf{m}_i \, \mathbf{p}_i \, \mathbf{v}_i]$ \label{CodIC}
		\State $\mathbf{h_i} \gets  \text{{HashInitialization}}(\,)$ \label{CodIH}
		\For{$j\gets0$ \textbf{until} $L_i-1$}\label{Loop1}
		\State $\mathbf{b}_{j} \gets \text{{MessageSplit}}(\mathbf{z}_i)$ \label{CodDM}
		\State $n \gets -1$
		\State $\mathbf{H}(n) \gets \text{{HashVariablesInitialization}}(\,)$ \label{CodIR}
		\For{$n\gets0$ \textbf{until} $79$}\label{Loop2}
        \State $\mathbf{w}(n) \gets \text{{WFunctionCalculation}}(n,\mathbf{b}_j)$ \label{CodCW}
        \State $\mathbf{f}(n) \gets \text{{FFunctionCalculation}}(n,\mathbf{B}(n),\mathbf{C}(n),\mathbf{D}(n))$ \label{CodCF}
		\State $\mathbf{H}(n) \gets \text{{HashVariablesUpdate}}(\mathbf{H}(n))$ \label{CodAVH}
		\EndFor \label{EndLoop2}
		\State $\mathbf{h}_i \gets  \text{{UpdateHash}}(\mathbf{H}(n))$ \label{CodAH}
		\EndFor \label{EndLoop1}
	\end{algorithmic}
\end{algorithm}

\subsection{Padding Insertion}

This step (lines \ref{CodGP} and \ref{CodIP} of the Algorithm \ref{ResumoSHA}) is performed before calculating the hash code and it makes the $i$-th message length, $m_i$, divisible by $M = 512$ after the Append Length step. The padding message, $p_i$, associated with the $i$-th incoming message is formed by a binary word of $ P_i $ bits in which the most significant bit is $ 1 $ and the rest of the bits are $ 0 $. The generation of the padding message is performed by the function PaddingGeneration($K_i$) shown in the line \ref{CodGP} of the algorithm \ref{ResumoSHA}. 

The calculation of the $P_i$ value can be expressed by
\begin{equation}
P_i = \left\{
\begin{array}{ll}
448 - (K_i \text{ mod }512) & \text{ for } (K_i \text{ mod }512) < 448 \\
512 - (K_i \text{ mod }512) + 448 & \text{ for }  (K_i \text{ mod }512) \geq 448
\end{array}
\right.,
\label{eq3}
\end{equation}
where the ($a$ $mod$ $b$) operation returns the \textit{modulo} of the division between $ a $ and $ b $. Thus, $p_i $ can be expressed as
\begin{equation}
\mathbf{}{p}_i = \left[
\begin{array}{cccc}
p_0 & p_1 & \dots & p_{P_i-1}
\end{array}
\right],
\label{eq4}
\end{equation}
where, $p_0=1$ and $p_i=0 \text{ for } i=1 \dots P_i-1$.

\subsection{Length Insertion}

In this step (lines \ref{CodGC} and \ref{CodIC} of the Algorithm \ref{ResumoSHA}) the message $v_i$ is added, which is characterized by a binary word of $T = 64 $ bits and expressed as
\begin{equation}
\mathbf{v}_i = \left[
\begin{array}{cccc}
v_0 & v_1 & \dots & v_{T-1}
\end{array}
\right] \text{ where } v_k \, \in \, \{0,1\} \, \forall \, k.
\label{eq5}
\end{equation}
The generation of the message length is performed by the function LenghtGeneration $(K_i)$ presented in the line \ref{CodGC} of the algorithm \ref{ResumoSHA}. The message $v_i$ stores the length value of the $i$-th incoming message $m_i$, that is, 
\begin{equation}
\mathbf{v}_i = \text{{Binary}}(K,T)
\label{eq6}
\end{equation}
where $Binary(a,b)$ is a function that returns a vector of size $b$ with the binary representation of a decimal number $a$ with $b$ bits according to the big-endian standard.

The 180-4 FIPS norm \cite{fips}, assumes that the size, $K_i$, of most messages can be represented by 64 bits, that is, $K_i<2^T$.

Finally, at the end of the second step the message, $z_i$, which is an extension of the $i$-th original input message, $m_i$, is generated (line \ref{CodIC} of the Algorithm \ref{ResumoSHA}). In this work the message $z_i$ is identified as a $Z_i$ bits vector expressed as
\begin{equation}
\mathbf{z}_i = \left[
\begin{array}{cccc}
z_0 & z_1 & \dots & z_{Z-1}
\end{array}
\right] \text{ where } z_k \, \in \, \{0,1\} \, \forall \, k.
\label{eq7}
\end{equation}

\subsection{Hash Code Initialization}
\label{iniciaH0-H4}

The hash code initialization (line \ref{CodIH} of the Algorithm \ref{ResumoSHA}) is standardized by the FIPS 180-4 \cite{fips} according to the following expressions:
\begin{equation}
\mathbf{ha}= \left[
\begin{array}{ccc}
h_{0}  & \dots & h_{31}
\end{array}
\right] = \text{{Binary}}(1732584193,32),
\label{eq9}
\end{equation}
\begin{equation}
\mathbf{hb}= \left[
\begin{array}{ccc}
h_{32}  & \dots & h_{63}
\end{array}
\right] = \text{{Binary}}(4023233417,32),
\label{eq10}
\end{equation}
\begin{equation}
\mathbf{hc}= \left[
\begin{array}{ccc}
h_{64}  & \dots & h_{95}
\end{array}
\right] = \text{{Binary}}(2562383102,32),
\label{eq11}
\end{equation}
\begin{equation}
\mathbf{hd}= \left[
\begin{array}{ccc}
h_{96}  & \dots & h_{127}
\end{array}
\right] = \text{{Binary}}(0271733878,32),
\label{eq12}
\end{equation}
and
\begin{equation}
\mathbf{he}= \left[
\begin{array}{ccc}
h_{128}  & \dots & h_{159}
\end{array}
\right] = \text{{Binary}}(3285377520,32),
\label{eq12}
\end{equation}
where 
\begin{equation}
\mathbf{h}_{i}= \left[
\begin{array}{ccccc}
\mathbf{ha}  & \mathbf{hb}&\mathbf{hc} & \mathbf{hd} & \mathbf{he}
\end{array}
\right].
\label{eq8}
\end{equation}

\subsection{Message Split}
\label{SudSec:DM}

In this step, line \ref{CodDM} of the Algorithm \ref{ResumoSHA}, the message $z_i$ is split into $L_i$ blocks of $M = 512$ bits, that is, 
\begin{equation}
\mathbf{z}_{i} = \left[
\begin{array}{cccc}
\mathbf{b}_{0} & \mathbf{b}_{1} & \dots & \mathbf{b}_{L_i-1}
\end{array}
\right],
\label{eq13}
\end{equation}
where each $j$-th block associated with $i$-th message is expressed as
\begin{equation}
\mathbf{b}_{j} = \left[
\begin{array}{cccc}
b_{j,0} & b_{j,1} & \dots & b_{j,M-1}
\end{array}
\right] \textnormal{ where } b_{j,k} \, \in \, \{0,1\} \, \forall \, k.
\label{eq14}
\end{equation}
The $j$-th block, $b_j$, can also be represented as
\begin{equation}
\mathbf{b}_{j} = \left[
\begin{array}{cccc}
\mathbf{u}_{j}[0] & \mathbf{u}_{j}[1] & \dots & \mathbf{u}_{j}[15]
\end{array}
\right],
\label{eqVetorB_j}
\end{equation}
where $\mathbf{u}_{j}[k]$ is a $ 32 $ bits message, that is,
\begin{equation}
\mathbf{u}_{j}[k] = \left[
\begin{array}{cccc}
u_{j}[k,0] & u_{j}[k,1] & \dots & b_{j}[k,31]
\end{array}
\right] 
\label{eq16}
\end{equation}
where $u_{j}[k,l] \, \in \, \{0,1\} \, \forall \, l$.

\subsection{$\mathbf{H}(n)$ Hash Variables Initialization }

The SHA-1 algorithm has five $ 32 $ bits variables, called $ A (n) $, $ B (n) $, $ C (n) $, $ D (n) $ and $ E (n) $ that are updated during iterations of the algorithm. These variables are identified in this work as vectors:
 
\begin{equation}
\mathbf{X}(n) = \left[
\begin{array}{cccc}
x_0 & x_1 & \dots & x_{31}
\end{array}
\right] \textnormal{ where } x_k \, \in \, \{0,1\} \, \forall \, k,
\label{eq17}
\end{equation}
where, the combination of these five variables form a vector of $ 160$ positions identified as
\begin{equation}
\mathbf{H}(n) = \left[
\begin{array}{ccccc}
\mathbf{A}(n) & \mathbf{B}(n) & \mathbf{C}(n) & \mathbf{D}(n) & \mathbf{E}(n)
\end{array}
\right].
\label{eq17a}
\end{equation}

The initialization of these variables, in the instant $ n = -1 $, (line \ref{CodIR} of the Algorithm \ref{ResumoSHA}) according to FIPS 180-4 \cite{fips} occurs with the receipt of the same values that start the hash $\mathbf{h}_i$, so $\mathbf{A}(-1) = \mathbf{ha}$, $\mathbf{B}(-1) = \mathbf{hb}$, $\mathbf{C}(-1) = \mathbf{hc}$, $\mathbf{D}(-1) =  \mathbf{hd}$ and $\mathbf{E}(-1) =  \mathbf{he}$.

\subsection{$\mathbf{w}(n)$ Variable Calculation }\label{SubW}

In SHA-1, it takes $80$ iterations for a valid output, $h_i$, associated with a $i$-th message be generated (Algorithm \ref{ResumoSHA}, line \ref{Loop2}). In each $n$-th iteration a $ w(n)$ variable is calculated, expressed as

\begin{equation}
\mathbf{w}(n) = \left\{
\begin{array}{ll}
\mathbf{u}_j[n] & \textnormal{ for } 0 \le n \le 15 \\
\mathbf{sw}[n] & \textnormal{ for }  16 \le n \le 79
\end{array}
\right.,
\label{eqWt}
\end{equation}
where

\begin{equation}
\mathbf{sw}[n] = \text{lr}\left(\mathbf{u}_j[n-3] \oplus \mathbf{u}_j[n-8]  \oplus \mathbf{u}_j[n-14] \oplus \mathbf{u}_j[n-16],1\right)
\label{eqWt1}
\end{equation}
where $\oplus$ is the exclusive or operation and lr$(r,s)$ represents the leftrotate function that is expressed as
\begin{equation}
\text{lr}(\mathbf{r},\mathbf{s}) = (\mathbf{r} \ll \mathbf{s}) \vee (\mathbf{r} \gg (32-\mathbf{s})),
\label{eqLR}
\end{equation}
where $\vee$, $\ll$, and $\gg$ are the bitwise OR and left and right bitwise shift, respectively.

\subsection{$f(\cdot)$ Function Calculation}\label{SubGF}

In each $n$-th iteration of each $j$-th block, $b_{j}(n)$, a nonlinear function, $f(\cdot)$, is calculated from the information of the hash variables $B(n)$, $C(n)$ and $D(n)$. The output of the function, $f(\cdot)$ is stored in the vector $f(n)$ (line \ref{CodCF} of the Algorithm \ref{ResumoSHA}), expressed as
\begin{equation}
\mathbf{f}(n)= f(n,\mathbf{B},\mathbf{C},\mathbf{D}) =\left\{
\begin{array}{l}
\alpha(n) \textnormal{ for } n=0 \dots 19 \\
\beta(n) \textnormal{ for } n=20 \dots 39 \\
\gamma(n) \textnormal{ for } n=40 \dots 59 \\
\delta(n) \textnormal{ for } n=60 \dots 79
\end{array}
\right.,
\label{eqF}
\end{equation}
where
\begin{equation}
\alpha(n) = (\mathbf{B}(n-1) \land \mathbf{C}(n-1)) \lor (\neg \mathbf{B}(n-1) \land \mathbf{D}(n-1)),
\label{eq20}
\end{equation}
\begin{equation}
\beta(n) =  \mathbf{B}(n-1) \oplus \mathbf{C}(n-1) \oplus \mathbf{D}(n-1),
\label{eq21}
\end{equation}
\begin{equation}
\gamma(n) = (\mathbf{B}(n-1) \wedge \mathbf{C}(n-1)) \vee ( \mathbf{B}(n-1) \wedge \mathbf{D}(n-1)) \vee ({C}(n-1) \wedge {D}(n-1))
\label{eq22}
\end{equation}
and
\begin{equation}
\delta(n) =  \mathbf{B}(n-1) \oplus \mathbf{C}(n-1) \oplus \mathbf{D}(n-1),
\label{eq23}
\end{equation}
where $\neg$ and $\wedge$ are negation operation and bitwise AND, respectively.

\subsection{Hash Variables Update}

Also, in each $n$-th iteration of each $j$-th block $b_{j}(n)$, the values of the variables $ A(n) $, $ B(n) $, $ C(n) $, $ D(n) $ and $ E(n) $ are updated after the calculation of $ f(n)$ (line \ref{CodAVH} of the Algorithm \ref{ResumoSHA}). The update of these variables is represented by the following equations:
\begin{equation}
\mathbf{E}(n) =  \mathbf{D}(n-1),
\label{eqEn}
\end{equation}
\begin{equation}
\mathbf{D}(n) =  \mathbf{C}(n-1),
\label{eqDn}
\end{equation}
\begin{equation}
\mathbf{C}(n) =  \textnormal{lr}(\mathbf{B}(n-1),30),
\label{eqCn}
\end{equation}
\begin{equation}
\mathbf{B}(n) =  \mathbf{A}(n-1)
\label{eqBn}
\end{equation}
and
\begin{equation}
\mathbf{A}(n) =  \mathbf{V}(n) + \mathbf{Z}(n) +  \text{lr}(A(n-1),5),
\label{eqAn}
\end{equation}
in which,
\begin{equation}
\mathbf{Z}(n)  =  \mathbf{W}(n) + \mathbf{E}(n-1)
\label{eqZn}
\end{equation}
and
\begin{equation}
\mathbf{V}(n)  =  \mathbf{f}(n) + \mathbf{k}(n).
\label{eqVn}
\end{equation}

The SHA-1 has four $ 32 $ bits constants $ k(n)$, which are used in the $n$-th iteration of each $j$-th block $b_j{n}$, as specified by
\begin{equation}
{K}(n)= \left\{
\begin{array}{ll}
1518500249 & \textnormal{ for } n=0 \dots 19 \\
1859775393 & \textnormal{ for } n=20 \dots 39 \\
2400959708 & \textnormal{ for } n=40 \dots 59 \\
3395469782 & \textnormal{ for } n=60 \dots 79
\end{array}
\right..
\label{eq24}
\end{equation}

\subsection{Hash Code Update}

For each $j$-th block, $b_{j}$, SHA-1 executes $ 80 $ iterations, and at the end of every $j$-th block the hash code is updated linearly following the expressions:
\begin{equation}
\mathbf{ha} = \mathbf{ha} + \mathbf{A}(79),
\label{eq33}
\end{equation}
\begin{equation}
\mathbf{hb} = \mathbf{hb} + \mathbf{B}(79),
\label{eq34}
\end{equation}
\begin{equation}
\mathbf{hc} = \mathbf{hc} + \mathbf{C}(79),
\label{eq35}
\end{equation}
\begin{equation}
\mathbf{hd} = \mathbf{hd} + \mathbf{D}(79),
\label{eq36}
\end{equation}
and
\begin{equation}
\mathbf{he} = \mathbf{he} + \mathbf{E}(79).
\label{eq37}
\end{equation}
So for every $ i $-th message, $m_i $, the value of the associated hash code, $ h_i $, is found in
\begin{equation}
N_i = L_i \times 80
\label{eqNi} 
\end{equation}
iterations, where $ N_i $ is defined in this work as the total number of interactions for the calculation of the hash associated with a message $m _i$.

\section{Proposed Implementation}\label{Sec:implementacao}

\begin{figure}[ht]
	\centering
	\includegraphics[width=1\textwidth]{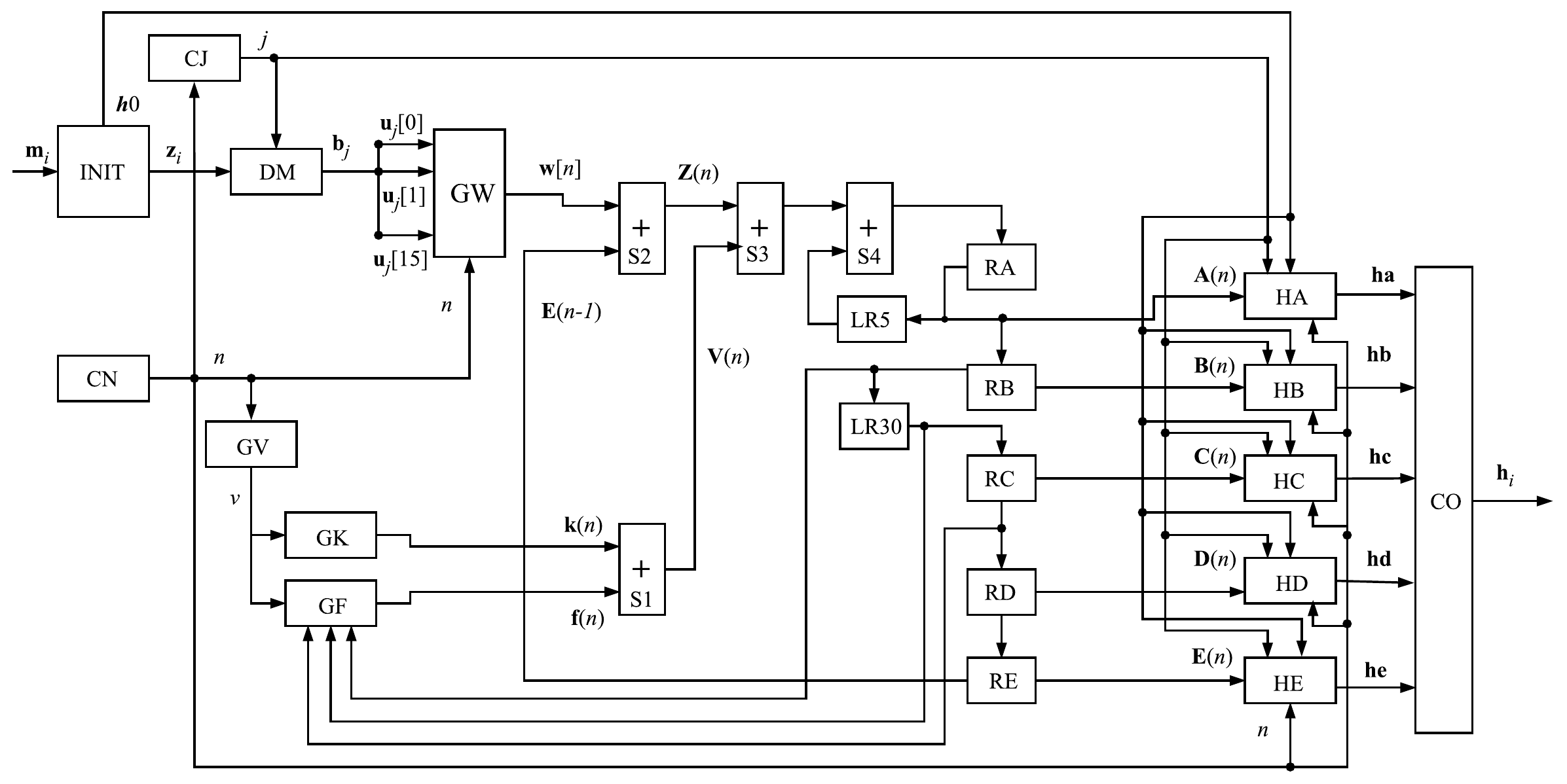}
	\caption{General architecture of the proposed SHA-1 hardware implementation}
	\label{SHA_Estrutura_Geral}
\end{figure}

Figure \ref{SHA_Estrutura_Geral} presents the general architecture of the proposed SHA-1 hardware implementation. The structure allows the visualization of the algorithm in Register Transfer Level (RTL), in which one can observe the signal flow (or variables) between the components of datapath and the registers $\textnormal{RA}$, $\textnormal{RB}$, $\textnormal{RC}$, $\textnormal{RD}$ and $\textnormal{RE}$. The hardware starts with the $i$-th message $m_i$ entering a module called $\textnormal{INIT}$ which is responsible for the functionalities presented between the lines \ref{Cod1} and \ref{CodIH} from the Algorithm \ref{ResumoSHA}, the control of the two loops (lines \ref{Loop1} and \ref{Loop2}) and the initialization of hash variables ($A(n)$, $B(n)$, $C(n)$, $D(n)$ and $E(n)$) to each $j$-th block, $b_j$, through the $ h0 $ signal.

The $\textnormal{CJ}$ and $\textnormal{CN}$ blocks are $\log_2(L)$ and $7$ bits counters, respectively. The $\textnormal{CN}$ counter is responsible for the loop iteration of line \ref{Loop2} of the Algorithm \ref{ResumoSHA}, generating the signal $n$. The $ \textnormal {CJ}$ counter is incremented by the counter $\textnormal {CN}$ and controls the loop iteration of the line \ref{Loop1} of the Algorithm \ref{ResumoSHA}, through the signal $j$. Based on line \ref{CodDM} of the Algorithm \ref{ResumoSHA} and subsection \ref{SudSec:DM}, the $ \textnormal{DM}$ module splits the $i$-th message $z_i$ into $L$ blocks of $ M = 512 $ bits, in which each $j$-th block is displayed in Figure \ref{SHA_Estrutura_Geral} by the signal $b_j$. This $M = 512 $ bits signal $b_j$ is then equally divided into $16$ buses of the $32$-bits, in which each $i$-th bus is represented by the signal $u_j[n]$. After this step, the $ w [n] $ signal  is generated by the $\textnormal {GW}$ module (line \ref{CodCW} of the Algorithm \ref{ResumoSHA}) from the signal counter  $ \textnormal{CN}$.
 
The modules $\textnormal{GF}$, $\textnormal{GK}$ and $\textnormal{GW}$ represent the operations expressed by Equations \ref{eqF}, \ref{eq24} and \ref{eqWt}, respectively. The modules $\textnormal{LR5}$ and $\textnormal{LR30}$ represent $\textnormal{leftrotate}$ operations expressed by Equations \ref{eqAn} e \ref{eqCn}. It is observed that unlike implementations in sequential processors such as GPP, uC (Micro-controllers) and others, these equations are executed in parallel, accelerating the SHA-1 algorithm. The details regarding the implementation of the modules  $\textnormal{GF}$, $\textnormal{GK}$, $\textnormal{GW}$, $\textnormal{LR5}$ and $\textnormal{LR30}$ are detailed in the following subsections.
 
 \subsection{ $\textnormal{GF}$ Module}\label{SecGF}
 
\begin{figure}[ht]
	\centering
	\includegraphics[width=80mm]{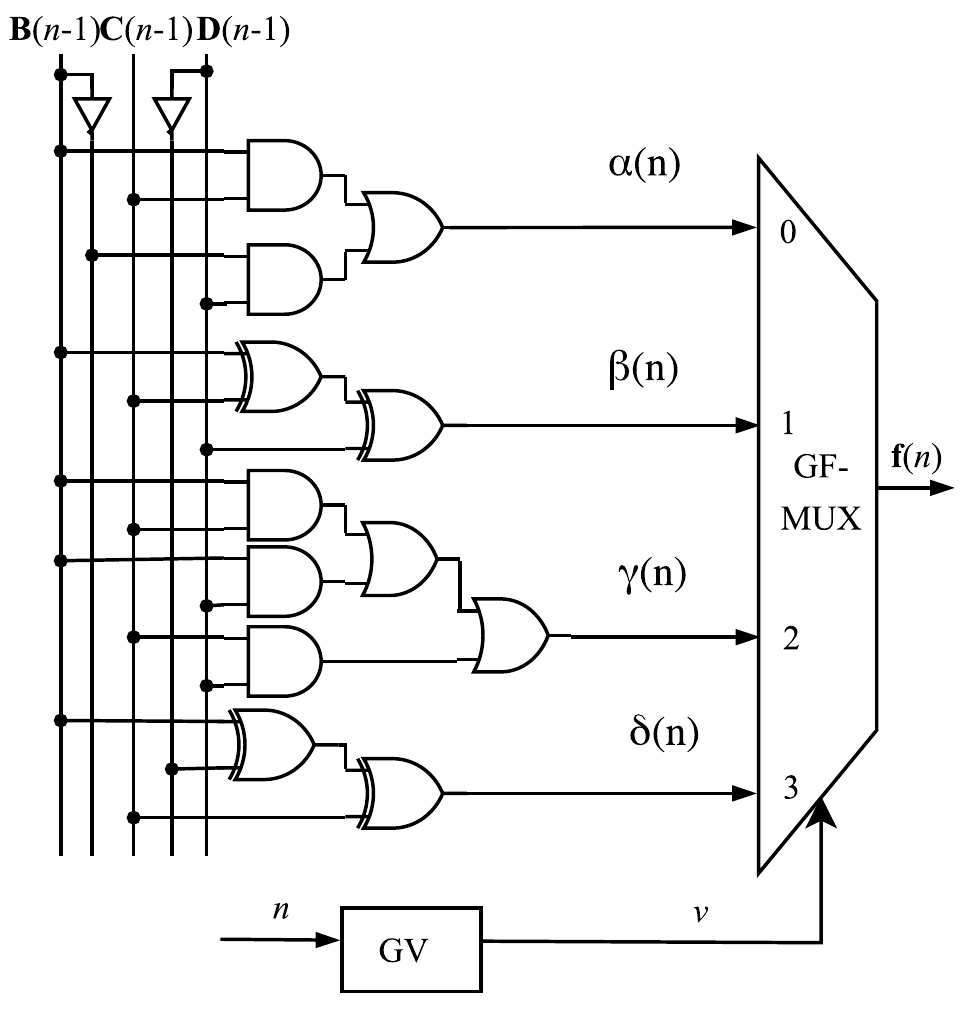}
	\caption{$\textnormal{GF}$ Module Architecture.}
	\label{FigGF}
\end{figure}

The $ \textnormal {GF}$ module implements the function described in subsection \ref{SubGF} and presented in the line \ref{CodCF} of the Algorithm \ref{ResumoSHA}. This module contains a multiplexer called $\textnormal {GF-MUX} $ which selects the function type from the $ n $ value according to Equation \ref{eqF} and detailed in Figure \ref{FigGF}.
 
 The function type selection in the $ \textnormal {GF-MUX} $ multiplexer is controlled by the $ \textnormal {GV}$ module, through binary logic with comparators and logic gates corresponding to each interval, having the following outputs,
 \begin{equation}
{GV}= \left\{
\begin{array}{ll}
0 & \textnormal{ for } n=0 \dots 19 \\
1 & \textnormal{ for } n=20 \dots 39 \\
2 & \textnormal{ for } n=40 \dots 59 \\
3 & \textnormal{ for } n=60 \dots 79
\end{array}
\right..
\label{eqGG}
\end{equation}
Each one selecting a function $ \mathbf {f}(n) $ based on the $7$ bits counter of the $ \textnormal {CN} $ module.

\subsection{$\textnormal{GW}$ Module}\label{SubSecW}

The $ \textnormal {GW} $ module consists of $ 16 $ messages $\mathbf{u}_{j}[n]$ (with $ 32 $-bits ) in the input, originating from $ \mathbf{b}_{j}[n]$, according to Equation \ref{eqVetorB_j} of Subsection \ref{SudSec:DM}, and has the purpose to perform the operation demonstrated by the equation \ref{eqWt} described in the subsection \ref{SubW} and line \ref{CodCW} of the Algorithm \ref{ResumoSHA}.

\begin{figure*}[h]
	\centering
	\includegraphics[width=1\textwidth]{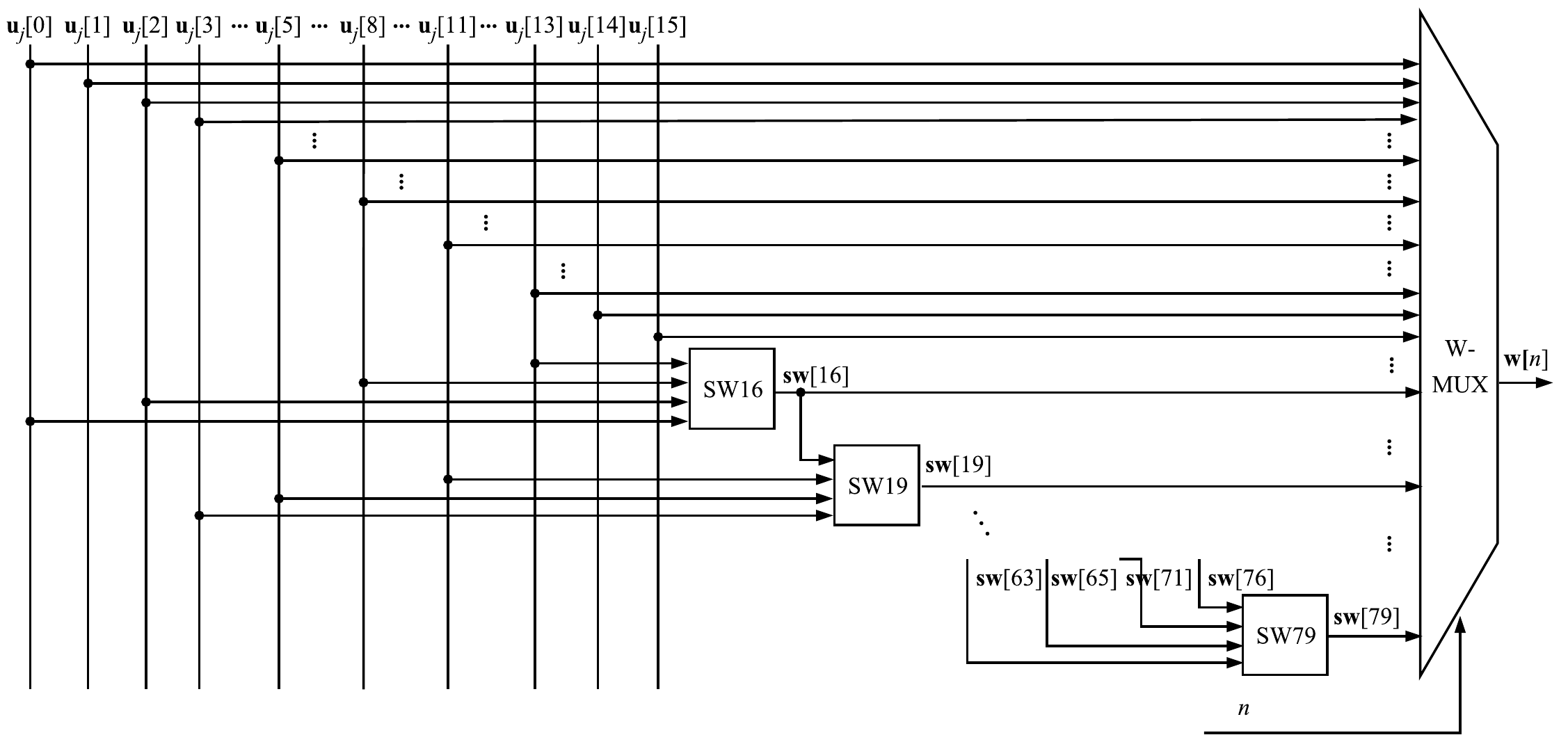}
	\caption{ $\textnormal{GW}$ Module Architecture}
	\label{fig:moduloW}
\end{figure*}

\begin{figure}[h]
	\centering
	\includegraphics[width=8.5cm,keepaspectratio]{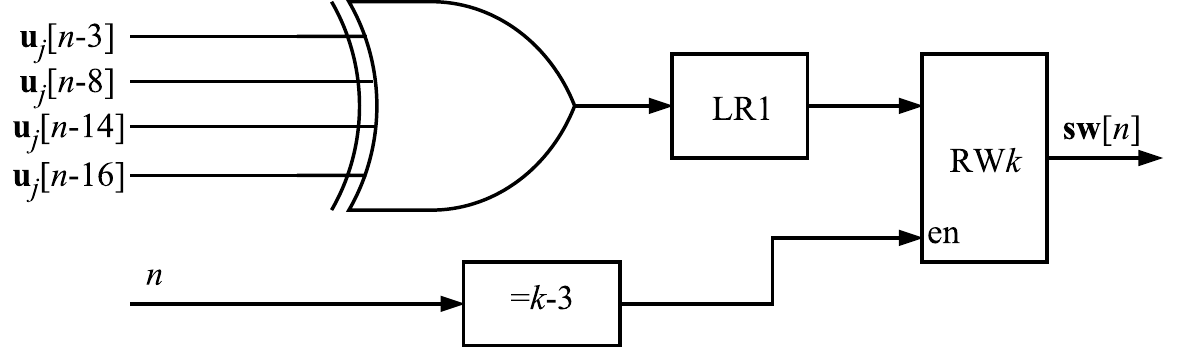}
	\caption{$\textnormal{SW}k$ Module Operation}
	\label{FiginsideW}
\end{figure}

Figure \ref{fig:moduloW} details the module that is formed by a $ 80 $ multiplexer inputs, called $\textnormal {W-MUX} $ which is selected from the $n$ signal. For the values of $ n $ from $ 16$ to $ 79 $, the signal $\mathbf{sw}[n]$ is expressed by the equation \ref{eqWt1} through the $ \text{SW} k $ module where $k=16\dots79$, specified in Figure \ref{FiginsideW}.

\begin{figure}[ht]
	\centering
	\includegraphics[width=75mm]{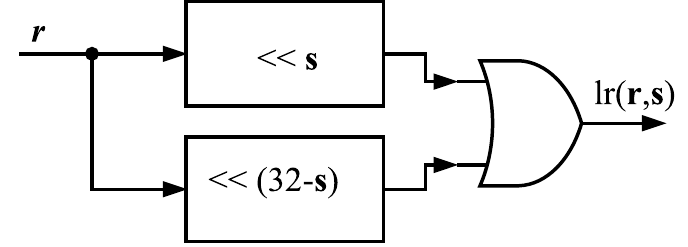}
	\caption{Arquitetura do módulo $\textnormal{RL}s$.}
	\label{FigRL}
\end{figure}

Each k-th module $\text{SW}k$ is formed by a register, called here $\text{RW}k$, a leftrotate module (Equation \ref{eqLR}) called LR1 a exclusive OR (XOR) gate and a comparator. The ${RW}k$ register stores the value of the signal $ \mathbf{sw}[n] $ through the comparator when $n=k-3$. The XOR logic gate performs the operation described in Equation \ref{eqWt1} and the LR1 module performs the leftrotate function for $\mathbf {s} = 1 $. Figure \ref{FigRL} details a generic module associated with the leftrotate implementation based on Equation \ref{eqLR}.

The $\textnormal {RL5} $ and $ \textnormal {RL30}$ modules perform the leftrotate operation for $ \mathbf {s} = $ 5 (see Equation \ref{eqAn}) and $\mathbf{s}=30$ (see Equation \ref{eqCn}), respectively. The implementation of these modules is also depicted by Figure \ref{FigRL}.

\subsection{$\mathbf{h}_i$ Hash Processing}

After generating the signals $\mathbf{w}(n)$, $\mathbf{k}(n)$, $\mathbf{f}(n)$, in each $ n$-th iteration, and the value $\mathbf{E}(n-1)$, the signals $\mathbf{Z}(n)$ and $\mathbf{V}(n)$, both of $ 32 $ bits, are calculated through the sum modules $\textnormal{S}1$ and $\textnormal{S}2$, executed in parallel, subsequently $\textnormal{S}3$ and  $\textnormal{S}4$. All the sum modules used in the implementation are $32$-bit-specific circuits, which optimize the processing time and the space occupied by the total circuit. The calculation of the signals $\mathbf{Z}(n)$ and $\mathbf{V}(n)$ are expressed by Equations \ref{eqZn} and \ref{eqVn} and are executed during the line \ref{CodAVH} of the Algorithm \ref{ResumoSHA}. The last step of every $n$-th iteration is the update of the hash variables, $\mathbf{A}(n)$, $\mathbf{B}(n)$, $\mathbf{C}(n)$, $\mathbf{D}(n)$ and $\mathbf{E}(n)$, stored in the $\textnormal{RA}$, $\textnormal{RB}$, $\textnormal{RC}$, $\textnormal{RD}$ e $\textnormal{RE}$ registers, respectively. Another important step in the SHA-1 is that one described in Section \ref{SubSecW}, in which the value of the $ \textnormal {RC}$ register is updated throught the $\textnormal{lr}(r,30)$ operation, expressed in detail by Equation \ref{eqLR}. At each iteration of $\mathbf {n} $ the values of the registers move between them by updating the other hash variables according to Equations \ref{eqEn} to \ref{eqAn}. These steps are executed in the line \ref{CodAVH} of the Algorithm \ref{ResumoSHA}.

At the end of the $ 80$ loop iterations in $ n $ (Line \ref{Loop2} of Algorithm \ref{ResumoSHA}), the parts that make up the hash, $\mathbf{ha}$, $\mathbf{hb}$, $\mathbf{hc}$, $\mathbf{hd}$ and $\mathbf{he}$ (Equation \ref{eq8}), are updated by the modules $\textnormal{HA}$, $\textnormal{HB}$, $\textnormal{HC}$, $\textnormal{HD}$ and  $\textnormal{HE}$, respectively. This step is performed on the line \ref{CodAH} of the Algorithm \ref{ResumoSHA}. Finally, at the end of the $ N_i$ iterations (Equation \ref{eqNi}) the hash code final value, $ \mathbf {h}_i $, associated to the $i-th$ message is achieved. The $ \textnormal {CO} $ module has the function of concatenating the $ 5 $ buses of the $ 32 $-bits formed by the signals  $\mathbf{ha}$, $\mathbf{hb}$, $\mathbf{hc}$, $\mathbf{hd}$ and $\mathbf{he}$ and generating a serial signal with the hash code $ \mathbf{h} _i $.

\section{Results}
\label{sec:resultados}

The Table \ref{Tab:Resultados} presents the results obtained after the hardware synthesis of the implementation proposed in this work (Figure \ref{SHA_Estrutura_Geral}). Results concerning the hardware occupancy in the target FPGA (Virtex 6 xc6vlx240t-11156) as well as the results associated with the latency and the throughput achieved after the synthesis process are presented. Results were generated for several parallel implementations of the SHA-1 algorithm according to Figure \ref{SHA_Estrutura_Geral}, differently from other works presented in \cite{Ref2SHA1,Ref3SHA1,Ref5SHA1,Ref6SHA1}, which used serial (pipeline) structures. The proposal here presented, used several SHA-1 parallel modules, enabling the throughout acceleration which is especially useful in cases of brute force password recovery, in which there are a large number of hash codes to be generated.

\begin{table}[ht]
	\begin{center}
		\caption{Results concerning occupancy, sampling rate and throughput for various parallel implementations of the SHA-1 algorithm.}
		\begin{tabular}{|c|c|c|c|c|c|c|}
			\hline
			$\textnormal{NI}$ &    $\textnormal{NR}$ & $\textnormal{PR}$ &$\textnormal{NLUT}$  & $\textnormal{PLUT}$ &$T_s$& $R_s$\\
			&      &       (\%) &   &(\%) &      ($\textnormal{ns}$) &  ($\textnormal{Gbps}$) \\
			\hline
			$1$ & $2.154$ & $0\textnormal{,}71$ & $2.605$ & $1\textnormal{,}72$ & $9\textnormal{,}932$ &  $0\textnormal{,}644$ \\
			\hline
			$4$ & $8.575$ & $2\textnormal{,}84$ & $10,388$ & $6\textnormal{,}89$ & $9\textnormal{,}961$ &  $2\textnormal{,}570$ \\
			\hline
			$8$ & $17.136$ & $5\textnormal{,}68$ & $20.662$ & $13\textnormal{,}71$ & $9\textnormal{,}965$ & $5\textnormal{,}138$  \\
		\hline
			$16$ & $34.255$ & $11\textnormal{,}36$ & $43.263$ &$28\textnormal{,}70$ &	$9\textnormal{,}994$ &$10\textnormal{,}246$ \\
			\hline
			$32$ & $68.498$ & $22\textnormal{,}72$ &$86.873$ &$57\textnormal{,}64$ &	$9\textnormal{,}994$ & $18\textnormal{,}296$ \\
			\hline
			$48$ & $102.733$&	$34\textnormal{,}08$ & $129.902$&$86\textnormal{,}19$ &	$10\textnormal{,}909$ & $28\textnormal{,}160$ \\
  			\hline            
		\end{tabular}
		\label{Tab:Resultados}
	\end{center}
\end{table}

The first column of the table, called $ \textnormal{NI} $, indicates the number of parallel implementations performed. The second column, $ \textnormal {NR} $, shows the number of registers used in the target FPGA and the third column, called $ \textnormal {PR} $, represents the percentage of registers used regarding the total available in the FPGA which is $ 301,440$. The fourth and fifth columns, called $ \textnormal {NLUT} $ and $ \textnormal {PLUT} $, represent the number of logical cells used as LUTs (Lookup Tables) for constructing digital circuits and the percentage of LUTs regarding the total amount available which is $ 150,720$. Finally, the sixth and seventh columns show the results, obtained for various implementations, of the sampling rate, $ T_s $ and throughput, $ R_s $, respectively.

The hardware, was developed in parallel as shown in Section \ref{Sec:implementacao}, with $ 32 $ bits buses so that the sampling time, $ T_s $, is corresponding to the clock, that is, every $n$-th iteration (see Loop of the line \ref{Loop2} of the Algorithm \ref{ResumoSHA}) is performed in a clock pulse time, here called $t_{\textnormal{CLK}}=T_s$. The $ T_s $ values are displayed in the sixth column of the table \ref{Tab:Resultados}. It is possible to verify that there is not a significant change with the increase of $ \textnormal {NI} $, that is, for an increment of $ 48\times$  of $ \textnormal {NI} $ there was only an increment of less than $ 1 $ ns in $ T_s $, which represents an increase of almost $ 32\times$  in hash throughput.

Based on the Algorithm \ref{ResumoSHA} and the architecture presented in Figure \ref{SHA_Estrutura_Geral}, for every $j$-th $M=512$ bits block $\mathbf{b}_j$, $ 80 $ iterations are executes (Equation \ref{eqNi}), so the proposed hardware throughput can be calculated as
\begin{equation}
R_s = \frac{ M \times \textnormal{NI} }{80 \times T_s} =\frac{512  \times \textnormal{NI}}{80 \times T_s}= \frac{64 \times \textnormal{NI}}{10 \times T_s}.
\label{eq46}
\end{equation}
It is important to note that the values of throughput greater than $15 \, \textnormal{Gbps}$ are unpublished in the literature ($\textnormal{NI}=32$ e $\textnormal{NI}=48$). A $ 28,16 Gbps $ throughput is equivalent to retrieve a totally unknown $ 6 $ digits numeric password (using the brute force method) in a maximum of $20 \, \textnormal{ms}$ or a $ 6 $ digits alpha numeric password (each digit with 62 possibilities) from a hash code in a maximum of $ 17.4 $ minutes.

\section{Conclusion}

This work presented a SHA-1 hardware implementation proposal. The proposed structure, also called ASSP, was synthesized in an FPGA aiming to validate the implemented circuit. All implementation details of the project were presented and analyzed regarding occupation area and processing time. The results obtained are quite significant and point to new possibilities of using hash algorithms in dedicated hardware for real-time and high-volume applications.

\section*{Funding}
This study was financed in part by the Coordena\c{c}\~ao de Aperfei\c{c}oamento de Pessoal de N\'ivel Superior (CAPES) - Finance Code 001.

\section*{Acknowledgments}
The authors wish to acknowledge the financial support of the Coordena\c{c}\~ao de Aperfei\c{c}oamento de Pessoal de N\'ivel Superior (CAPES) for their financial support.

\bibliography{PaperMain}

\begin{thebibliography}{19}
\expandafter\ifx\csname natexlab\endcsname\relax\def\natexlab#1{#1}\fi
\providecommand{\url}[1]{\texttt{#1}}
\providecommand{\href}[2]{#2}
\providecommand{\path}[1]{#1}
\providecommand{\DOIprefix}{doi:}
\providecommand{\ArXivprefix}{arXiv:}
\providecommand{\URLprefix}{URL: }
\providecommand{\Pubmedprefix}{pmid:}
\providecommand{\doi}[1]{\href{http://dx.doi.org/#1}{\path{#1}}}
\providecommand{\Pubmed}[1]{\href{pmid:#1}{\path{#1}}}
\providecommand{\bibinfo}[2]{#2}
\ifx\xfnm\relax \def\xfnm[#1]{\unskip,\space#1}\fi
\bibitem[{Al-Kiswany et~al.(2009)Al-Kiswany, Gharaibeh, Santos-Neto \&
  Ripeanu}]{Ref2GPUSHA1}
\bibinfo{author}{Al-Kiswany, S.}, \bibinfo{author}{Gharaibeh, A.},
  \bibinfo{author}{Santos-Neto, E.}, \& \bibinfo{author}{Ripeanu, M.}
  (\bibinfo{year}{2009}).
\newblock \bibinfo{title}{On gpu's viability as a middleware accelerator}.
\newblock {\it \bibinfo{journal}{Cluster Computing}\/},  {\it
  \bibinfo{volume}{12}\/}, \bibinfo{pages}{123--140}.
\bibitem[{Iyer \& Mandal(2013)}]{Ref4SHA1}
\bibinfo{author}{Iyer, N.~C.}, \& \bibinfo{author}{Mandal, S.}
  (\bibinfo{year}{2013}).
\newblock \bibinfo{title}{Implementation of secure hash algorithm-1 using
  fpga}.
\newblock {\it \bibinfo{journal}{International Journal of Information and
  Computation Technology}\/},  {\it \bibinfo{volume}{3}\/},
  \bibinfo{pages}{757--764}. \URLprefix
  \url{https://www.ripublication.com/irph/ijict_spl/04_ijictv3n8spl.pdf}.
\bibitem[{Jarvinen(2004)}]{Ref1SHA1}
\bibinfo{author}{Jarvinen, K.} (\bibinfo{year}{2004}).
\newblock \bibinfo{title}{Design and implementation of a sha-1 hash module on
  fpgas}, .
\newblock \URLprefix
  \url{http://cwcserv.ucsd.edu/\~billlin/classes/ECE111/SHA1\-Javinen.pdf}.
\bibitem[{Kakarountas et~al.(2006)Kakarountas, Michail, Milidonis, Goutis \&
  Theodoridis}]{Ref3SHA1}
\bibinfo{author}{Kakarountas, A.~P.}, \bibinfo{author}{Michail, H.},
  \bibinfo{author}{Milidonis, A.}, \bibinfo{author}{Goutis, C.~E.}, \&
  \bibinfo{author}{Theodoridis, G.} (\bibinfo{year}{2006}).
\newblock \bibinfo{title}{High-speed fpga implementation of secure hash
  algorithm for ipsec and vpn applications}.
\newblock {\it \bibinfo{journal}{The Journal of Supercomputing}\/},  {\it
  \bibinfo{volume}{37}\/}, \bibinfo{pages}{179--195}. \URLprefix
  \url{http://dx.doi.org/10.1007/s11227-006-5682-5}.
  \DOIprefix\doi{10.1007/s11227-006-5682-5}.
\bibitem[{Kara et~al.(2017)Kara, Alistarh, Alonso, Mutlu \& Zhang}]{Ref11}
\bibinfo{author}{Kara, K.}, \bibinfo{author}{Alistarh, D.},
  \bibinfo{author}{Alonso, G.}, \bibinfo{author}{Mutlu, O.}, \&
  \bibinfo{author}{Zhang, C.} (\bibinfo{year}{2017}).
\newblock \bibinfo{title}{Fpga-accelerated dense linear machine learning: A
  precision-convergence trade-off}.
\newblock In {\it \bibinfo{booktitle}{2017 IEEE 25th Annual International
  Symposium on Field-Programmable Custom Computing Machines (FCCM)}\/} (pp.
  \bibinfo{pages}{160--167}).
\newblock \DOIprefix\doi{10.1109/FCCM.2017.39}.
\bibitem[{Khan et~al.(2014)Khan, ul~Abideen \& Paracha}]{Ref5SHA1}
\bibinfo{author}{Khan, S.}, \bibinfo{author}{ul~Abideen, Z.}, \&
  \bibinfo{author}{Paracha, S.~S.} (\bibinfo{year}{2014}).
\newblock \bibinfo{title}{An ultra low power and high throughput fpga
  implementation of sha-1 hash algorithm}.
\newblock {\it \bibinfo{journal}{International Journal of Computer Science and
  Information Security}\/},  {\it \bibinfo{volume}{12}\/},
  \bibinfo{pages}{80--86}. \URLprefix
  \url{http://sites.google.com/site/ijcsis/}.
\bibitem[{Lee et~al.(2009)Lee, Lee, Park \& Cho}]{Ref3aSHA1}
\bibinfo{author}{Lee, E.-H.}, \bibinfo{author}{Lee, J.-H.},
  \bibinfo{author}{Park, I.-H.}, \& \bibinfo{author}{Cho, K.-R.}
  (\bibinfo{year}{2009}).
\newblock \bibinfo{title}{Implementation of high-speed sha-1 architecture}.
\newblock {\it \bibinfo{journal}{IEICE Electronics Express}\/},  {\it
  \bibinfo{volume}{6}\/}, \bibinfo{pages}{1174--1179}.
\bibitem[{Marks \& Niewiadomska{-}Szynkiewicz(2014)}]{Ref1GPUSHA1}
\bibinfo{author}{Marks, M.}, \& \bibinfo{author}{Niewiadomska{-}Szynkiewicz,
  E.} (\bibinfo{year}{2014}).
\newblock \bibinfo{title}{Hybrid {CPU/GPU} platform for high performance
  computing}.
\newblock In {\it \bibinfo{booktitle}{28th European Conference on Modelling and
  Simulation, {ECMS} 2014, Brescia, Italy, May 27-30, 2014}\/} (pp.
  \bibinfo{pages}{508--514}).
\bibitem[{Michail et~al.(2014)Michail, Athanasiou, Theodoridis \&
  Goutis}]{Ref7SHA1}
\bibinfo{author}{Michail, H.}, \bibinfo{author}{Athanasiou, G.},
  \bibinfo{author}{Theodoridis, G.}, \& \bibinfo{author}{Goutis, C.}
  (\bibinfo{year}{2014}).
\newblock \bibinfo{title}{On the development of high-throughput and
  area-efficient multi-mode cryptographic hash designs in fpgas}.
\newblock {\it \bibinfo{journal}{Integration, the VLSI Journal}\/},  {\it
  \bibinfo{volume}{47}\/}, \bibinfo{pages}{387 -- 407}.
\bibitem[{Michail et~al.(2005)Michail, Kakarountas, Koufopavlou \&
  Goutis}]{Ref2SHA1}
\bibinfo{author}{Michail, H.}, \bibinfo{author}{Kakarountas, A.~P.},
  \bibinfo{author}{Koufopavlou, O.}, \& \bibinfo{author}{Goutis, C.~E.}
  (\bibinfo{year}{2005}).
\newblock \bibinfo{title}{A low-power and high-throughput implementation of the
  sha-1 hash function}.
\newblock In {\it \bibinfo{booktitle}{2005 IEEE International Symposium on
  Circuits and Systems}\/} (pp. \bibinfo{pages}{4086--4089 Vol. 4}).
\bibitem[{Michail et~al.(2016)Michail, Athanasiou, Theodoridis, Gregoriades \&
  Goutis}]{Ref6SHA1}
\bibinfo{author}{Michail, H.~E.}, \bibinfo{author}{Athanasiou, G.~S.},
  \bibinfo{author}{Theodoridis, G.}, \bibinfo{author}{Gregoriades, A.}, \&
  \bibinfo{author}{Goutis, C.~E.} (\bibinfo{year}{2016}).
\newblock \bibinfo{title}{Design and implementation of totally-self checking
  sha-1 and sha-256 hash functions' architectures}.
\newblock {\it \bibinfo{journal}{Microprocessors and Microsystems}\/},  {\it
  \bibinfo{volume}{45}\/}, \bibinfo{pages}{227 -- 240}.
\bibitem[{{Network Working Group}(2001)}]{rfc}
\bibinfo{author}{{Network Working Group}} (\bibinfo{year}{2001}).
\newblock \bibinfo{title}{{Request for Comments: 3174}}.
\newblock \bibinfo{howpublished}{http://www.faqs.org/rfcs/rfc3174.html}.
\bibitem[{{NIST}(2013)}]{dss:13}
\bibinfo{author}{{NIST}} (\bibinfo{year}{2013}).
\newblock \bibinfo{title}{Digital signature standard (dss)}.
\newblock {\it \bibinfo{journal}{FIPS PUB 186-4}\/}, . \URLprefix
  \url{http://nvlpubs.nist.gov/nistpubs/FIPS/NIST.FIPS.186-4.pdf}.
\bibitem[{{NIST}(2015)}]{fips}
\bibinfo{author}{{NIST}} (\bibinfo{year}{2015}).
\newblock \bibinfo{title}{{Secure Hash Standard (SHS)}}.
\newblock
  \bibinfo{howpublished}{http://nvlpubs.nist.gov/nistpubs/FIPS/NIST.FIPS.180-4.pdf}.
\bibitem[{Shaikh et~al.(2017)Shaikh, Kalwar, Memon \& Sheikh}]{Ref12}
\bibinfo{author}{Shaikh, F.}, \bibinfo{author}{Kalwar, I.~H.},
  \bibinfo{author}{Memon, T.~D.}, \& \bibinfo{author}{Sheikh, S.}
  (\bibinfo{year}{2017}).
\newblock \bibinfo{title}{Design and analysis of linear phase fir filter in
  fpga using pso algorithm}.
\newblock In {\it \bibinfo{booktitle}{2017 6th Mediterranean Conference on
  Embedded Computing (MECO)}\/} (pp. \bibinfo{pages}{1--4}).
\newblock \DOIprefix\doi{10.1109/MECO.2017.7977216}.
\bibitem[{Shi et~al.(2012)Shi, Ma, Cote \& Wang}]{RefLivro1}
\bibinfo{author}{Shi, Z.}, \bibinfo{author}{Ma, C.}, \bibinfo{author}{Cote,
  J.}, \& \bibinfo{author}{Wang, B.} (\bibinfo{year}{2012}).
\newblock \bibinfo{title}{Hardware implementation of hash functions}.
\newblock In {\it \bibinfo{booktitle}{Introduction to Hardware Security and
  Trust}\/} (pp. \bibinfo{pages}{27--50}).
\newblock \bibinfo{publisher}{Springer}.
\bibitem[{de~Souza \& Fernandes(2014)}]{Ref10}
\bibinfo{author}{de~Souza, A.}, \& \bibinfo{author}{Fernandes, M.}
  (\bibinfo{year}{2014}).
\newblock \bibinfo{title}{Parallel fixed point implementation of a radial basis
  function network in an fpga}.
\newblock {\it \bibinfo{journal}{Sensors}\/},  {\it \bibinfo{volume}{14}\/},
  \bibinfo{pages}{18223--18243}.
\bibitem[{Stallings(2015)}]{stallings:15}
\bibinfo{author}{Stallings, W.} (\bibinfo{year}{2015}).
\newblock {\it \bibinfo{title}{Criptografia e seguran\c{c}\~a de redes}\/}.
\newblock (\bibinfo{edition}{6th} ed.).
\newblock \bibinfo{publisher}{Pearson Education do Brasil}.
\bibitem[{Venkateshan et~al.(2015)Venkateshan, Patel \& Varghese}]{Ref13}
\bibinfo{author}{Venkateshan, S.}, \bibinfo{author}{Patel, A.}, \&
  \bibinfo{author}{Varghese, K.} (\bibinfo{year}{2015}).
\newblock \bibinfo{title}{Hybrid working set algorithm for svm learning with a
  kernel coprocessor on fpga}.
\newblock {\it \bibinfo{journal}{IEEE Transactions on Very Large Scale
  Integration (VLSI) Systems}\/},  {\it \bibinfo{volume}{23}\/},
  \bibinfo{pages}{2221--2232}. \DOIprefix\doi{10.1109/TVLSI.2014.2361254}.

\end{thebibliography}

\end{document}